\documentclass[twocolumn,showpacs,preprintnumbers,amsmath,amssymb]{revtex4}
\usepackage{graphicx}
\usepackage{dcolumn}
\usepackage{bm}
\usepackage{epsfig}
\newcommand{\be}{\begin{equation}}
\newcommand{\ee}{\end{equation}}
\newcommand{\ba}{\begin{eqnarray}}
\newcommand{\ea}{\end{eqnarray}}
\newcommand{\bi}{\begin{itemize}}
\newcommand{\ei}{\end{itemize}}

\newcommand{\ovl}{\overline}

\newcommand{\half}{{\textstyle\frac{1}{2}}}

\newcommand{\quarter}{{\textstyle\frac{1}{4}}}

\newcommand{\<}{\langle}
\renewcommand{\>}{\rangle}
\newcommand{\eq}{Eq.~}

\newcommand{\fig}{Fig.~}

\newcommand{\la}{\label}

\newcommand{\txts}{\textstyle}

\begin{document}

\preprint{MIT-CTP 3881}

\title{A calculation of the bulk viscosity in SU(3) gluodynamics} 

\author{Harvey~B.~Meyer}
\email{meyerh@mit.edu}
\affiliation{Center for Theoretical Physics\\ 
       Massachusetts Institute of Technology\\
                 Cambridge, MA 02139, U.S.A.}

\date{\today}

\begin{abstract}
We perform a lattice Monte-Carlo calculation of the trace-anomaly two-point function
at finite temperature in the SU(3) gauge theory. 
We obtain the long-distance properties of the correlator in the continuum limit
and extract the bulk viscosity  $\zeta$ via a Kubo formula. Unlike the tensor correlator
relevant to the shear viscosity, the scalar correlator depends strongly on temperature.
If $s$ is the entropy density, we find that $\zeta/s$ becomes rapidly small at high $T$,
 $\zeta/s<0.15$ at $1.65T_c$ and $\zeta/s<0.015$ at $3.2T_c$.
However $\zeta/s$ rises dramatically just above $T_c$, with $0.5<\zeta/s<2.0$ at $1.02T_c$.
\end{abstract}

\maketitle

Given the success of the hydrodynamical description~\cite{huovinen} 
of high-energy heavy ion reactions, it is of primary interest to compute the 
shear and bulk viscosities of the quark-gluon plasma in the region 
$T_c\leq T\leq 3T_c$ relevant to the RHIC and the forthcoming LHC experiments.

Phenomenology combined with hydrodynamics calculations
point to a very small shear viscosity to entropy density ratio, 
$\eta/s<0.2$~\cite{teaney}, 
and viscous calculations find even smaller values~\cite{paulrom}.
The main uncertainty in these determinations is the sensitivity 
to the initial conditions of the hydro-regime.
A recent lattice calculation of $\eta/s$ 
in the purely gluonic plasma~\cite{hm-shear} found values 
in the range 0.1--0.2, under an assumption that we shall return to below. 
These values are significantly smaller
than the leading-order perturbative result $\eta/s\approx0.8$
for $\alpha_s=0.3$~\cite{arnold-shear}.

In this Letter we carry out a lattice calculation of the bulk viscosity, $\zeta$,
in the SU(3) gauge theory, reusing much of the technology developed 
in~\cite{hm-shear}. The calculation relies on a Kubo formula and 
on the reconstruction of the real-time 
spectral function~\cite{karsch-visco,huang93,nakamura}
from the Euclidean trace-anomaly correlator.
This is known to be a numerically hard problem, in particular some assumptions
have to be made on the spectral function in order to extract the viscosity
--- typically, an assumption of smoothness.
In this respect, the bulk viscosity is a more favorable case than the 
shear viscosity at high temperatures, because the spectral function 
is rigorously known to be smooth at $T\to\infty$. 
This is related to the fact that the bulk viscosity
vanishes for any conformal field theory, and non-Abelian gauge theories become
conformal at high energies by becoming free.

It has recently been argued~\cite{kharzeev}, based on an exact sum rule, lattice data
on $\epsilon-3P$ and a simple model for the spectral function, 
that the bulk viscosity rises sharply close to the (first order) phase transition.
This is remarkable, because a gas of structureless point particles has
negligible bulk viscosity both in the non-relativistic and the extreme-relativistic
limits~\cite{weinberg71}, and  the bulk viscosity is often neglected for that reason 
even when the shear viscosity is taken into account.
Furthermore, the leading-order perturbative QCD result~\cite{arnold-bulk} for 
$\zeta/s$ is numerically very small: to within $10\%$ for $0.06<\alpha_s<0.3$,
\be
\zeta/s = 0.020\alpha_s^2~~~~~(N_{\rm f}=0).
\la{eq:LO}
\ee
It  satisfies the relation~\cite{weinberg71}
$\zeta\approx 15\eta[{\txts\frac{1}{3}}-v_s^2]^2$,
$v_s$ being the velocity of sound in the medium.
Instead, AdS/CFT calculations that incorporate deviations from conformality 
at strong coupling find  $\zeta\propto\eta[{\txts\frac{1}{3}}-v_s^2] $~\cite{buchel}. 
These formulas suggest that the dynamics responsible for the bulk viscosity 
away from conformality, $v_s^2\!<\!\frac{1}{3}$, is different at strong and weak coupling.

We will show  that $\zeta/s$ depends strongly on temperature: it is 
phenomenologically negligible above $2T_c$, while it becomes ${\rm O}(1)$ 
just above $T_c$. We will also acquire some knowledge of the spectral 
function from first principles thay may help interpret this striking behavior.

\section*{Methodology}
The explicit form of the Euclidean energy-momentum tensor $T_{\mu\nu}$
can be found for instance in~\cite{gluex};
we write $T_{00}= \ovl T_{00}+\quarter\theta$, where $\theta$ is the trace anomaly.
With $L_0=1/T$ the inverse temperature, we consider the two-point functions
($0<x_0<L_0$)
\ba
C_\theta(x_0) & =&  L_0^5{\txts\int} d^3{\bf x} ~\< \theta(0)\theta(x_0,{\bf x}) \> \\
C_{\bar0}(x_0) &=& L_0^5{\txts\int} d^3{\bf x} ~\<\ovl T_{00}(0)\ovl T_{00}(x_0,{\bf x}) \>,\\
C_{\bar0\theta}(x_0) &=& L_0^5 {\txts\int} d^3{\bf x} ~\< \ovl T_{00}(0)\theta(x_0,{\bf x}) \>.
\la{eq:Cx0}
\ea
This correlator is represented by a spectral function,
\be
C_\theta(x_0) = L_0^5 \int_0^\infty \rho_\theta(\omega) 
\frac{\cosh \omega(\half L_0-x_0)}{\sinh \frac{\omega L_0}{2}} d\omega,
\la{eq:C=int_rho}
\ee
in terms of which the bulk viscosity is given by~\cite{jeon}
\be
\zeta(T) = \frac{\pi}{9} \left.\frac{d\rho_\theta}{d\omega}\right|_{\omega=0}.
\ee
Important properties of $\rho_\theta$ are its positivity, $\rho_\theta(\omega)/\omega\geq0$
and parity, $\rho_\theta(-\omega)=-\rho_\theta(\omega)$.
At tree level, 
 $C^{\rm t.l.}_\theta(x_0) = \frac{484d_A}{9\pi^4}\alpha_s^2N^2
 [f(\tau)-\frac{\pi^4}{60}]$,
with $\tau$, $d_A$ and $f(\tau)$ as in~\cite{hm-shear}, and
\ba
\rho_\theta^{\rm t.l.}(\omega) &=& 
 \frac{A_{\rm t.l.} ~\omega^4 }{\tanh\quarter \omega L_0},~~  \la{eq:tl}
A_{\rm t.l.} = d_A \left(\frac{11\alpha_sN}{3(4\pi)^2}\right)^2.
\ea
By contrast with the shear viscosity case~\cite{hm-shear}, 
the absence of a $\delta$-function at the origin corresponds to the fact 
that the theory is conformal at tree-level and hence the bulk viscosity vanishes.

Because $\int d^3{\bf x} \<T_{00}(x) {\cal O}(0)\> = T^2\partial_T \<{\cal O}\>_T$
for any local operator ${\cal O}$ and $x_0\neq 0$, the two other correlators are related 
to $C_\theta$ by 
\ba
C_{\bar{0}}(x_0) &=& {\txts\frac{1}{16}} C_\theta(x_0) +b_{\bar0}(T),\la{eq:C0}\\
C_{\bar0\theta}(x_0) &=& - {\txts\frac{1}{4}} C_\theta(x_0) + b_\theta(T),\la{eq:C0th}\\
b_{\bar 0}(T)&\equiv& \half T^{-3}\partial_T(\epsilon-3P) + 3 T^{-3}s, \la{eq:b0}\\
b_\theta(T) &\equiv&   T^{-3} \partial_T (\epsilon-3P). \la{eq:bth}
\ea
i.e. they only differ by an $x_0$-independent shift.
In perturbation theory, $b_\theta(T) = \frac{11d_A}{54}(N\alpha_s)^2 
+{\rm O(\alpha_s^3)}$ and $b_{\bar 0}(T)$ is also easily extracted 
from~\cite{Kajantie:2002wa}.
This completes the perturbative result 
for the Euclidean correlators up to ${\rm O}(\alpha_s^2)$.

We finally introduce the following moments of the spectral function ($n=0,1,\dots$):
\be
 \<\omega^{2n}\>\equiv L_0^5\int_0^\infty d\omega 
   \frac{\omega^{2n}\rho_\theta(\omega)}{\sinh \omega L_0/2}
       = \left.\frac{d^{2n}C_\theta}{dx_0^{2n}}\right|_{x_0=L_0/2}
\ee
The latter equality implies that they are directly accessible to  lattice calculations.
For a spectral function that grows like $\omega^4$ at large frequencies, 
it is natural to quote the pure numbers $\frac{4!}{4^n(4+2n)!}\<\omega^{2n}\>/T^{2n}$.

\begin{figure}[t]
\begin{center}
\psfig{file=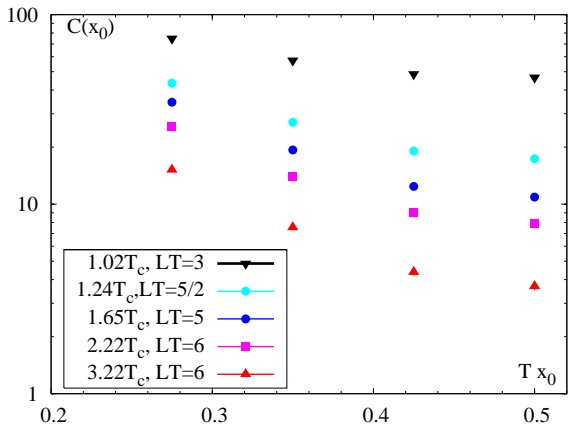,angle=-90,width=8.5cm}
\end{center}
\caption{The correlator $C_\theta(x_0)$ on $L_0/a=8$ lattices.}
\la{fig:Cbulk-tlimp}
\end{figure}

\section*{Numerical results}
We simulate the isotropic Wilson action with a two-level algorithm~\cite{hm-ymills}, 
as in~\cite{hm-shear}, 
and employ the `bare-plaquette' discretization of $\ovl T_{00}$ and $\theta$ as 
in~\cite{gluex}. We use the data of~\cite{necco-sommer} to 
determine the value of $Tr_0$ and to compute the lattice beta-function,
and use $T_cr_0=0.75(1)$~\cite{boyd96} to obtain $T/T_c$.

We present the $C_\theta$ correlator, computed on $L_0/a=8$ lattices
in the temperature range $T_c<T<3T_c$, in \fig\ref{fig:Cbulk-tlimp}. 
It has been `improved' by the same technique as in~\cite{hm-shear}, 
so that cutoff effects are ${\rm O}(g_0^2a^2)$. The deviations from 
conformality are significantly larger than those seen in the 
tensor channel relevant to the shear viscosity~\cite{hm-shear}.
As $T$ approaches $T_c$ from above, they become very large.

We investigated finite-volume effects at $1.65$ and $1.24T_c$ 
by varying $LT$ from 3.5 to 5 and 2.5 to 3.67 respectively;
no statistically significant dependence was observed.

\begin{figure}[t]
\begin{center}
\psfig{file=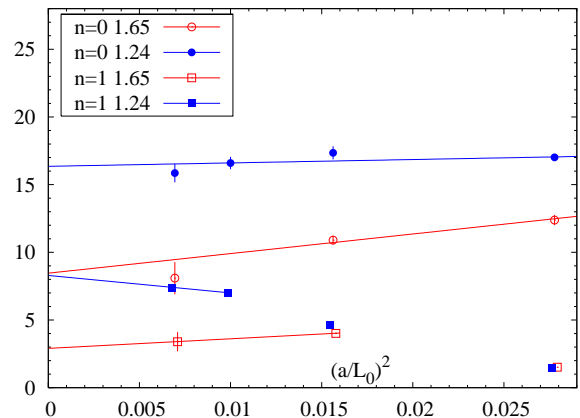,angle=-90,width=8.5cm}
\end{center}
\caption{Continuum extrapolation of the first two moments $\<\omega^0\>$ 
and $\<\omega^2\>/(120T^2)$ of $\rho_\theta(\omega,T)$ for $T=1.24$ and 
$1.65T_c$.  The $\<\omega^0\>$  continuum limits are resp.
$16.3(5)$ and $8.5(8)$.}
\la{fig:extrapol}
\end{figure}
Figure \ref{fig:extrapol} shows the continuum extrapolation of 
the first two moments of $\rho_\theta$ from lattices $L_0/a=6,8,10,12$.
The scaling region for  $\<\omega^0\>$ starts at $L_0/a=6$, while for
$\<\omega^2\>$ it presumably only starts at $L_0/a=8$ or 10.
We also find $s/T^3=4.46(6)$ and $5.48(7)$ at $T=1.24$ and $1.65T_c$ in the continuum.

On the lattice, the relations (\ref{eq:C0}--\ref{eq:bth}) are only satisfied 
up to O($a^2$) corrections. They can be used to over-constrain the continuum
extrapolation of the $C_\theta$ correlator. It turns out that the 
direct measurement of $C_\theta$ is  more accurate than the estimators based on 
relations (\ref{eq:C0},\ref{eq:C0th}) by at least a factor five, 
and that $C_{\bar0}$ has large discretization errors; 
we have nonetheless checked that the data is consistent with 
relations (\ref{eq:C0}--\ref{eq:bth}) in the continuum limit.
While the spectral representation~(\ref{eq:C=int_rho}) is in principle only valid 
in that limit, smaller lattice spacings will be necessary to
extrapolate the whole correlator to the continuum. 

We thus proceed to reconstruct the spectral function at the smallest lattice spacing, 
based on $N$ points of the correlator at abscissas $\{x_0^{(i)}>2a\}$, with
(strictly positive-definite) covariance matrix $S_{ii'}$.
We write~\cite{hm-shear} $\rho(\omega) = m(\omega)~[1+a(\omega) ]$  
where $m(\omega)\!>\!0$ has the high-frequency \mbox{behavior} of~\eq\ref{eq:tl}. 
In practice we have used 
$m(\omega)= A_{\rm t.l.}\omega^4/[\tanh(\quarter\omega L_0)\tanh^2(c \omega L_0)]$ 
where $c$ is a tunable parameter 
typically set to $\quarter$. We build an estimator of $a(\omega)$,
$ \widehat a(\omega) = {\txts\sum_{\ell=1}^N} c_\ell u_\ell(\omega)$,
using the same basis  $\{u_\ell\}$ as in~\cite{hm-shear}.
Upon discretization of the $\omega$ variable, $M(x_0,\omega)\equiv  K(x_0,\omega) m(\omega)$
 becomes an $N\times N_\omega$ matrix
$M_{ij}$, and the functions $\{u_\ell\}_{\ell=1}^{N}$ are the columns of the matrix $U$
in the singular-value decomposition of  $M^{\rm t}=U{\rm w}V^{\rm t}$.
The positivity of $\widehat\rho(\omega)=m(\omega)[1+\widehat a(\omega)]$ 
provides an important \emph{a posteriori} consistency check. 
The spectral functions obtained in this way from $L_0/a=12$ lattices
are shown on \fig\ref{fig:sf}, for which $N=4$.

\begin{figure}[t]
\begin{center}
\psfig{file=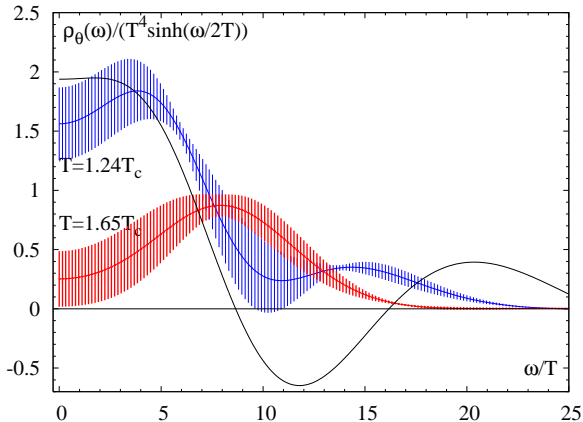,angle=-90,width=8.5cm}
\end{center}
\caption{The result for $\rho_\theta(\omega)$ from $L_0/a=12$ lattices. 
The bulk viscosity is given by $\zeta/T^3=(\pi/18)\times {\rm intercept}$.
The oscillating curve is the (rescaled) resolution function $\widehat\delta(0,\omega)$.}
\la{fig:sf}
\end{figure}

Solving for the coefficients $c_\ell$ results in
$\widehat a(\omega)=\sum_{i=1}^N \Delta C_i q_i(\omega)$,
with  $\Delta C_i\doteq C(x_0^{(i)})-\int d\omega M_i(\omega)$ 
and $q_i(\omega)=V_{i\ell}{\rm w}_\ell^{-1} u_\ell(\omega)$,
which is related to the genuine spectral function by 
\ba
\widehat a(\omega) &=& {\txts\int_0^\infty} 
 d\omega'~\widehat\delta(\omega,\omega')~a(\omega'),\\
\widehat\delta(\omega,\omega') &=& {\txts\sum_{i=1}^N}~ q_i(\omega)~ M_i(\omega') .
\ea
The \emph{resolution function} $\widehat\delta(0,\omega) $ is plotted in \fig\ref{fig:sf}.
Clearly it has the effect of smoothing $a(\omega)$ over a region 
$\Delta\omega\approx 5T$ around the origin; note that 
$a(\omega)\stackrel{\omega\to\infty}{\to}0$.

In solving inverse problems, it is often found necessary, when $N$ becomes
large, to introduce a `regulator',
whose role is to prevent the reconstruction from becoming unstable, at the cost
of making the resolution function $\widehat\delta(\omega,\omega')$
less sharply peaked around $\omega'=\omega$. One way~\cite{bg} to implement this 
in our method is to replace the $q_i(\omega)$ by 
${[{\rm w}V^{\rm t}+\lambda V^{\rm t}S]^{-1}}_{i\ell}~u_\ell(\omega)$,
where $\lambda$ is typically chosen so that $\widehat\rho$ yields a 
$\chi^2/{\rm d.o.f.}$ of order unity. We have however not found it
necessary to make  $\lambda\neq0$ in any of the cases presented here.

The value of $\widehat\rho/\omega|_{\omega=0}$ and its statistical error, 
\emph{together with the resolution function},
form an assumption-free representation of the information on $\zeta$ 
contained in the lattice data. 
In order to quote a systematic error on $\zeta$, 
some modelling of the spectral function is necessary.

Take for instance $T=1.24T_c$, 
where our best estimate is $\widehat\zeta/s=0.065(17)_{\rm stat}$.
Increasing $c$ from $\quarter$ to $\half$ reduces $\widehat\zeta/s$ to $0.010(2)$
but causes $\widehat\rho$ to assume negative values around $\omega=10T$ by 
one standard deviation. One may therefore regard this as a lower bound on $\zeta/s$.
We also infer a conservative upper bound based on the positivity of 
$\rho$~\cite{hm-shear}. Setting $\widehat\rho=\rho_{\rm t.l.}$ for
$\omega>20T_c$, and introducing a Breit-Wigner of width $2T$ at the origin, 
we find that $\zeta/s<0.37$ at $90\%$ confidence level
in order for that contribution to the correlator not to exceed $C_\theta(L_0/2)$.
We obtain similarly $\zeta/s<0.15$ at $1.65T_c$.

\begin{figure}[t]
\begin{center}
\psfig{file=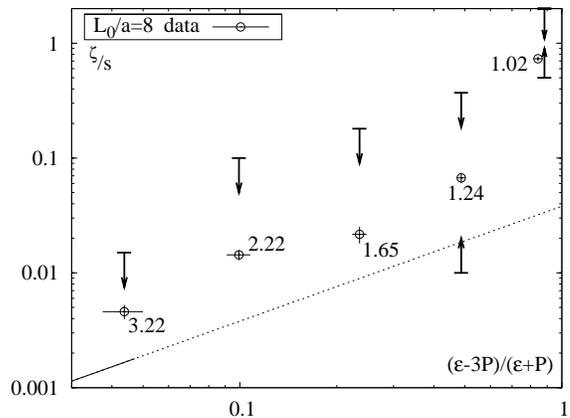,angle=-90,width=8.5cm}
\end{center}
\caption{The  bulk viscosity in units of the entropy density,
 as a function of the conformality measure
$\frac{\epsilon-3P}{\epsilon+P}$. The statistical errors are shown,
as well as the bounds explained in the text.
The solid line is the perturbative prediction \eq\ref{eq:LO}, naively
continued beyon $\alpha_s=0.3$ by the dashed line.}
\la{fig:zeta}
\end{figure}
\section*{Bulk viscosity \& non-conformality}
 It is clear from~\fig\ref{fig:sf} that for $T\stackrel{>}{\to}T_c$
the spectral function increases overall and also the spectral weight clusters
around the origin. Both features contribute to an increase in the bulk viscosity,
and they are robust predictions, since they essentially correspond to
the increase of $\<w^0\>$ and the decrease of $\<w^2\>/(T^2\<w^0\>)$.
We now study the relation of this effect to non-conformality in more detail.
As a measure of (non)-conformality, we use $\frac{\epsilon-3P}{\epsilon+P}$,
which is easier to compute than the velocity of sound but to which it is related 
at weak coupling by
\[1-3v_s^2 ={\txts\frac{4}{3}}(\epsilon-3P)/(\epsilon+P)~[1+{\rm O}(\alpha_s)]. \]
We obtain $\epsilon-3P$ from~\cite{boyd96} by quadratic interpolation
and compute $\epsilon+P$ ourselves. In \fig\ref{fig:zeta}, we show 
the ratio $\zeta/s$ obtained on $L_0/a=8$ lattices, where we have 
data at five different temperatures, as a function of this measure.
We  define a renormalized, ``$(\epsilon-3P)$-scheme coupling'', via 
$\frac{\epsilon-3P}{\epsilon+P}\stackrel{.}{=}\frac{55N^2\alpha_s^2(2\pi T)}{96\pi^2}$,
and insert it into \eq\ref{eq:LO}. That leads to the curve shown on the 
same plot.  It lies generally below the lattice points and, unsurprisingly,
does not account for the sharp rise of $\zeta$ close to $T_c$.


%
\section*{Conclusion}
%
We have computed the correlation functions of the energy-momentum tensor 
relevant to the bulk viscosity in the SU(3) pure gauge theory,
as well as the entropy density $s$, to high accuracy.
Our best estimates of the bulk viscosity, obtained on $L_0/a=12$ lattices, are
\be
 \zeta/s = \left\{ \begin{array}{l@{~~~}l}
  0.008(7)\Big[\begin{array}{c}\scriptstyle{0.15}\\ \scriptstyle{0} \end{array}\Big]
& (T=1.65T_c,~LT={\txts\frac{16}{3}}) \\
  0.065(17)\Big[\begin{array}{c}\scriptstyle{0.37}\\ \scriptstyle{0.01}\end{array}\Big]
 & (T=1.24T_c,~LT={\txts\frac{8}{3}}).
           \end{array} \right.
\la{eq:zeta}
\ee
where the statistical error is given and the square bracket specifies 
conservative upper and lower bounds.
This constitutes the first successful lattice calculation of the bulk viscosity 
of an SU($N$) gauge theory. In \eq\ref{eq:zeta}, the bounds hold \emph{in the continuum}.
Since we now have estimates of both $\eta$~\cite{hm-shear} and $\zeta$ at $1.24T_c$, we can 
ask whether the relation $\zeta\approx 15\eta[{\txts\frac{1}{3}}-v_s^2]^2$
holds. Using $v_s^2\approx0.25$~\cite{boyd96}, the formula predicts $\zeta/\eta\approx0.1$,
while our central value is $0.6$. Our data thus points to a much larger bulk viscosity than
the formula suggests.
A non-negligible bulk viscosity is a signature of the interplay between translational
and internal degrees of freedom; a small shear viscosity that of a strongly interacting system.

The bulk viscosity is non-vanishing below $1.3T_c$,
where deviations from conformality in $\epsilon$ and $P$ are large.
In particular we  observe a sharp rise in $\zeta$ in an $L_0/a=8$ simulation 
just above the first order phase transition,
\be
\zeta/s = 0.73(3) 
 \Big[\begin{array}{c}\scriptstyle{2.0}\\\scriptstyle{0.5}\end{array}\Big] 
\qquad   (T=1.02T_c,~LT=3).
\ee
Looking to `ordinary' substances, 
it is worth noting  that the bulk viscosity of nitrogen remains finite on 
the liquid-vapor coexistence line~\cite{eu}, while it becomes very large
close to the endpoint (for the case of xenon see~\cite{moldover}).
Although QCD with physical $u,d,s$ quark masses seems to undergo a crossover rather
than a phase transition~\cite{fodor},
the rise of $\zeta$ is related to the large rate of increase of $(\epsilon-3P)/T^4$
\cite{kharzeev} and to the clustering of the spectral weight towards $\omega=0$ (\fig\ref{fig:sf}).
We therefore also expect
an increase in the bulk viscosity around 200MeV in the real world. 
The effect could manifest itself in heavy ion
collisions by an enhanced entropy production in the later stages of the reaction,
provided the viscous hydrodynamical description is still valid at that time.



I am grateful to Krishna Rajagopal and John Negele for 
their encouragement and stimulating discussions. I further thank G.~Aarts,
M.~Alford, J.-W.~Chen, Ph.~de~Forcrand, D.~Kharzeev, P. Petreczky, R. Venugopalan
for useful discussions. The more difficult simulations
were performed on the BlueGeneL at M.I.T., and I thank Andrew Pochinsky for 
his assistance in running it successfully. This work was supported in part by 
funds provided by the U.S. Department of Energy under cooperative research agreement
DE-FG02-94ER40818.



\begin{thebibliography}{99}

\bibitem{huovinen}
  P.~F.~Kolb, P.~Huovinen, U.~W.~Heinz and H.~Heiselberg,
  Phys.\ Lett.\  B {\bf 500}, 232 (2001);
  P.~Huovinen, P.~F.~Kolb, U.~W.~Heinz, P.~V.~Ruuskanen and S.~A.~Voloshin,
  Phys.\ Lett.\  B {\bf 503}, 58 (2001);
  D.~Teaney, J.~Lauret and E.~V.~Shuryak,
  Phys.\ Rev.\ Lett.\  {\bf 86}, 4783 (2001).

\bibitem{teaney}
  D.~Teaney,
  Phys.\ Rev.\  C {\bf 68}, 034913 (2003).

\bibitem{paulrom}
  P.~Romatschke and U.~Romatschke,
  arXiv:0706.1522 [nucl-th];
  H.~Song and U.~W.~Heinz,
  arXiv:0709.0742.

\bibitem{hm-shear}
  H.~B.~Meyer,
  arXiv:0704.1801 [hep-lat], Rapid Comm., in press.

 \bibitem{arnold-shear}
   P.~Arnold, G.D.~Moore and L.G.~Yaffe,
   JHEP {\bf 0305}, 051 (2003).

\bibitem{karsch-visco}
  F.~Karsch and H.W.~Wyld,
  Phys.\ Rev.\  D {\bf 35}, 2518 (1987).

\bibitem{huang93}
  S.~Huang,
  Phys.\ Rev.\  D {\bf 47}, 653 (1993).

\bibitem{nakamura}
  A.~Nakamura and S.~Sakai,
  Phys.\ Rev.\ Lett.\  {\bf 94}, 072305 (2005).


\bibitem{kharzeev}
  D.~Kharzeev and K.~Tuchin,
  arXiv:0705.4280 [hep-ph].

\bibitem{weinberg71}
  S.~Weinberg,
  Astrophys.\ J.\  {\bf 168}, 175 (1971).

\bibitem{arnold-bulk}
  P.~Arnold, C.~Dogan and G.~D.~Moore,
  Phys.\ Rev.\  D {\bf 74}, 085021 (2006)
  [arXiv:hep-ph/0608012].

\bibitem{buchel}
  P.~Benincasa, A.~Buchel and A.~O.~Starinets,
  Nucl.\ Phys.\  B {\bf 733}, 160 (2006)
  [arXiv:hep-th/0507026];
  A.~Buchel,
  arXiv:0708.3459 [hep-th].











\bibitem{necco-sommer}
  S.~Necco and R.~Sommer,
  Nucl.\ Phys.\ B {\bf 622} (2002) 328.

\bibitem{hm-ymills}
  H.B.~Meyer,
  JHEP {\bf 0401}, 030 (2004).


\bibitem{jeon}
  S.~Jeon and L.~G.~Yaffe,
  Phys.\ Rev.\  D {\bf 53}, 5799 (1996)
  [arXiv:hep-ph/9512263].


\bibitem{boyd96}
  G.~Boyd, J.~Engels, F.~Karsch, E.~Laermann, C.~Legeland, M.~Lutgemeier and B.~Petersson,
  Nucl.\ Phys.\  B {\bf 469}, 419 (1996).


\bibitem{Kajantie:2002wa}
  K.~Kajantie, M.~Laine, K.~Rummukainen and Y.~Schroder,
  Phys.\ Rev.\  D {\bf 67}, 105008 (2003)
  [arXiv:hep-ph/0211321].

\bibitem{gluex}
  H.~B.~Meyer and J.~W.~Negele,
  arXiv:0707.3225 [hep-lat].
{The trace anomaly $\theta$ is denoted by $S^{\rm g}$ there.}


\bibitem{fodor}
  Y.~Aoki, G.~Endrodi, Z.~Fodor, S.~D.~Katz and K.~K.~Szabo,
  Nature {\bf 443}, 675 (2006)
  [arXiv:hep-lat/0611014];
  M.~Cheng {\it et al.},
  arXiv:0710.0354 [hep-lat].

\bibitem{bg}
G.E.~Backus, F.~Gilbert, Geophys. J. of the Royal Astronomical Society, 
{\bf 16}, 169 (1968); Philosoph. Transact. of the Royal Society of London A, 
{\bf 266}, 123 (1970).


\bibitem{eu}  
K.~Rah and B.V.~Eu, 
J.~Chem.~Phys. {\bf 114},23 10436 (2001)

\bibitem{moldover} 
K.A.~Gillis, I.I.~Shindler, M.R.~Moldover, 
Phys.\ Rev.\  E {\bf 72}, 051201 (2005).

\end{thebibliography}
\end{document}